\newcommand{\doublespace}{\addtolength{\baselineskip}{0.5\baselineskip}}

\documentstyle[12pt,epsf]{article}

\begin{document}
\doublespace

\title{Deciphering Secure
Chaotic Communication}
%A Simple Technique for Decoding an Unknown
% Modulated Chaotic
%Time-series
%}

\author{
C. Mathiazhagan\thanks{e-mail address: mathi@ee.iitm.ernet.in}
\\
    Department of Electrical Engineering
\\
    Indian Institute of Technology, Madras
\\
  Chennai 600 036, India}

\maketitle
%\doublespace
\begin{abstract}

  A simple technique for decoding an unknown modulated chaotic time-series
 is presented.
 We point out that, by fitting a polynomial model to the modulated
chaotic
  signal, the error in the fit gives sufficient
information
  to decode the modulating signal.
 For analog implementation, a lowpass filter can be
 used for fitting.
 This method is simple and easy to
implement
 in hardware.

\end{abstract}

 Indexing terms: chaotic time-series, secure communication
 (PACS: 05.45).

\vspace{.2in}

  Secure communication using chaos has received much attention recently.
 Various methods for modulating and demodulating a chaotic oscillator
 have been proposed (see \cite{opp} and references cited in \cite{corron}).
 There are two ways to hide the modulating signal inside the
 chaotic carrier signal. One way is to add a small modulating
 signal to the chaotic carrier signal whose amplitude is
 much larger.
 This is called
 chaotic masking. In this case, power is wasted in the carrier.
 Another way to encode the modulating signal is to modulate the parameters
 of the carrier chaotic oscillator. This method has been demonstrated
 experimentally by many groups\cite{corron}.

  However, it has also been pointed out that secure communication using
 chaos can be broken (\cite{perez},\cite{short}). The first method in \cite{perez}
 is specific to the Lorenz oscillator case. The second method by Short\cite{short}
 is more general. In this approach by Short, the modulating signal is
 assumed to be small and the phase-space of the carrier chaotic oscillator
 is reconstructed from the transmitted time-series using the standard
 delay embedding techniques (\cite{farmer}, \cite{brynmawr}). The chaotic
 time-series is predicted by noting the flow of nearby trajectories in the
 embedded phase-space. Then the fourier transform of the difference between
 the predicted series and the actual transmitted series is taken and a comb
filter
  is applied. This fourier spectrum now reveals the modulating signal.
 As noted in \cite{vp}, this technique works well when the modulating
 signal amplitude is small. If the amplitude of the modulating signal
 is large, the phase-space structure of the carrier gets
greatly altered and it may not be possible to get a good
delay embedding of the carrier oscillator dynamics\cite{vp}.

  Another way to unmask the modulating signal is to plot
 correlation integral\cite{brynmawr}
 of the chaotic time-series as a function of time.
This also contains some information about the
 modulating signal. However, this method requires lot of computation and
 works only when the modulating waveform is very slowly varying.

  In this brief, we point out a much simpler way to extract the modulating
 signal without using the phase-space information or sophosticated
 frequency filtering. This method is simple and easy to implement in
 hardware for real-time decoding and works well when the
 modulating parameter variation is sufficiently large.
 This method has been tested for many
 continous-time chaotic oscillator systems, including the Glass-Mackey
 equation that generates more complex chaotic waveforms.

%%%%%%%%%%%%%%%%%%%%%%%%%%%%%%%%%%%%%%%%%%
% include appropriate directory for figs.
%%%%%%%%%%%%%%%%%%%%%%%%%%%%%%%%%%%%%%%%%%
\begin{figure}[h]
  \hspace{0.7in}
  \epsffile{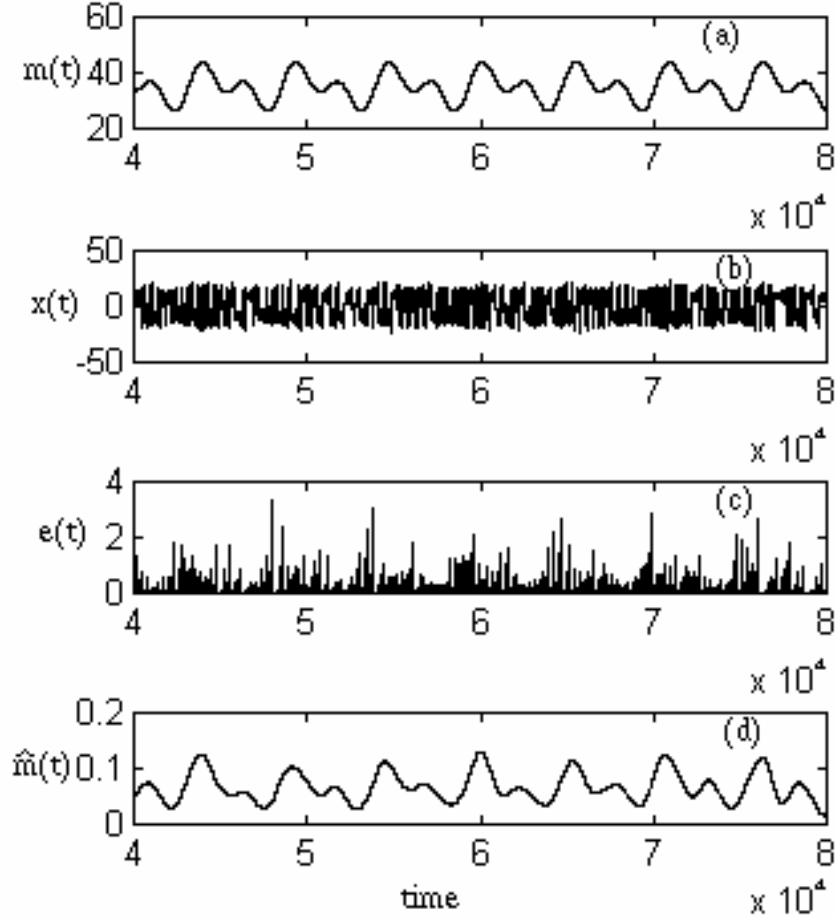}
  \caption{ Results for the Lorenz oscillator case. (a) shows
            the modulating waveform, (b) the transmitted variable
            $x(t)$, (c) the fit error signal and (d) demodulated
            output $\hat{m}(t)$.  }
 \end{figure}

  We take a short segment of the data and fit a polynomial.
 The number of data points for the fit is kept more than the order of
 the polynomial.
 The error in the fit now gives some information about the modulating
 waveform. As the modulationg parameter is varied, the Lyapunov exponents
 ($\lambda$'s) of
 the chaotic oscillator vary and the waveform also varies from, say, a
 `more' chaotic nature to `less' chaotic nature. When the waveform is
 less chaotic (smaller positive $\lambda$), we expect the error in the
 fit to be smaller. The error in the fit becomes larger when the modulating
 parameter moves the oscillator to a more chaotic region (larger positive
 $\lambda$).

 Since the information about the initial fit is lost within
 a time interval of about $1/\lambda$, next fit to a nearby segment
 gives another independent estimate of the fit error.
 It is easy to get
 sufficient number of such error points for averaging. The fit error data
 is then averaged (for about 100-200 points) and applied to a lowpass
 filter. We used a simple 4th order Butterworth lowpass filter.
 In case of a single-tone modulating signal,
 accurate demodulation is easy and a sharp bandpass
filter will do. In a more realistic communication system, the
 baseband modulating signal will consist of a band of frequencies.
  The results presented here are
 shown for a two-tone modulating signal of the type
 $m(t) = A_{1} \sin \omega_{1} t + A_{2} \sin \omega_{2} t$.

  We have explored other minor variations of the above technique
 for calculating the fit error. Some improvement in the demodulated
 waveform
 can be found by averging the demodulated signals of different
 orders of polynomial curve fit. Another way to find the fit error
 is to project the fitted polynomial to a nearby time location and
 calculate the fit error in that location. The improvements are
 only marginal.

  First, we show the results for the Lorenz oscillator case with parameters
 $\sigma =10$, $b=8/3$ and $r=35$ (notations as in \cite{corron}).
 For modulation, the parameter
  $r$ is varied around its nominal value by the
two-tone signal with
  amplitudes $A_{1}=A_{2}=5$ and $\omega_{1} =
 2\omega_{2} = 0.233$.
  That is:
$r=r_{0}+A_{1}\sin \omega_{1}t +A_{2}\sin \omega_{2}t $.
  The integration is
done using a 4th order Runge-Kutta method
  with a time step of $\Delta
T=0.01$. The variable $x(t)$ of the Lorenz
 oscillator  is assumed to be transmitted.
 A 4th order polynomial is used
 to fit every 10 points in the time-series.
 A 4th order lowpass filter with a small cut-off frequency
 (0.001/$\Delta T$) is used
 filter the error signal.
 The results are summarised in Fig. 1.
 The demodulated output $\hat m(t)$ is shown in Fig. 1d. The fourier
frequency
  spectrum of the error signal (before the final lowpass filter)
 is shown in Fig. 2. It is clear that the
 fit error data contains information about the modulating waveform.

 \begin{figure}[h]
  \hspace{0.5in}
  \epsffile{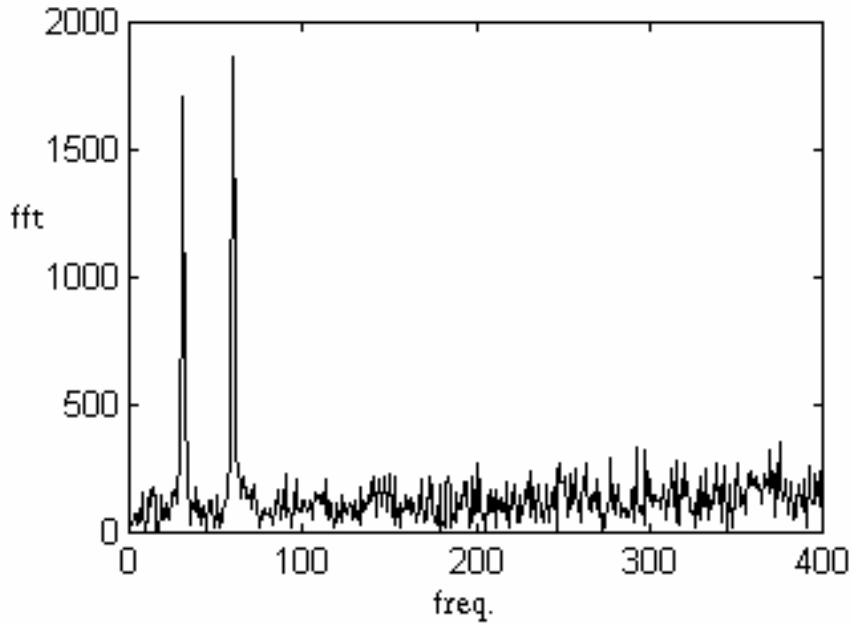}
  \caption{ The fourier spectrum of fit error signal $e(t)$. }
 \end{figure}

 This demodulation technique requires
 a digital signal processor (DSP) for real-time decoding. To aviod this,
 we present another simpler method that can be easily implemented
 with analog circuitry (Fig. 3). Here, instead of using a DSP
 or a computer,
 we use a simple lowpass filter LPF1 (of order 1 or 2)
 with low time-constant to predict a short segment of the input chaotic
 time-series. Then the error $e(t)$ between the input signal $x(t)$
 and the LPF1
output
  is calculated. The error can be computed with a squaring or an
absolute value
  circuitry.
 The delay in the input path can be implemented with
 a simple I or II order allpass filter.
 This is to compensate for the delay in LPF1.
 LPF2 is a sharper lowpass filter (say, of order 4 or 5)
 with a large time-constant that averages the error signal and produces the
 output demodulated signal $\hat m(t)$. This analog implementation also
gives
  performance comparable to the polynomial fit algorithm given earlier.

 \begin{figure}[h]
  \hspace{0.75in}
  \epsffile{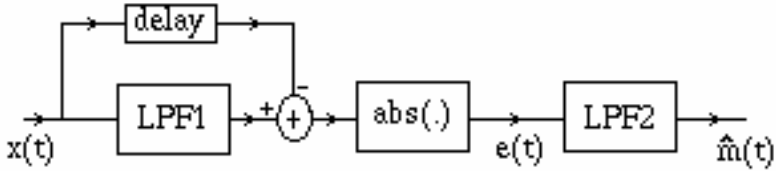}
  \caption{ The analog implementation. }
 \end{figure}

 The results for the Rossler oscillator case
are shown Fig. 4.
  The Rossler oscillator is simulated with parameters
$a=0.2$, $b=0.2$ and
  $c=5.7$. The $b$ parameter is modulated, as before,
with a two-tone modulating
  signal $A_{1}=A_{2}=0.015$ and $\omega_{1} =
 2\omega_{2} = 0.02$. The time-step
  for Runge-Kutta method is $\Delta
T=0.15$. LPF1 is a 2nd order Butterwoth lowpass filter
  with a large cut-off
frequency
 ($0.15/\Delta T$).  The averaging lowpass filter
  LPF2 is a 4th
order
 Butterwoth filter with a small cut-off frequency ($0.002/\Delta T$).

 \begin{figure}[h]
  \hspace{.25in}
  \epsffile{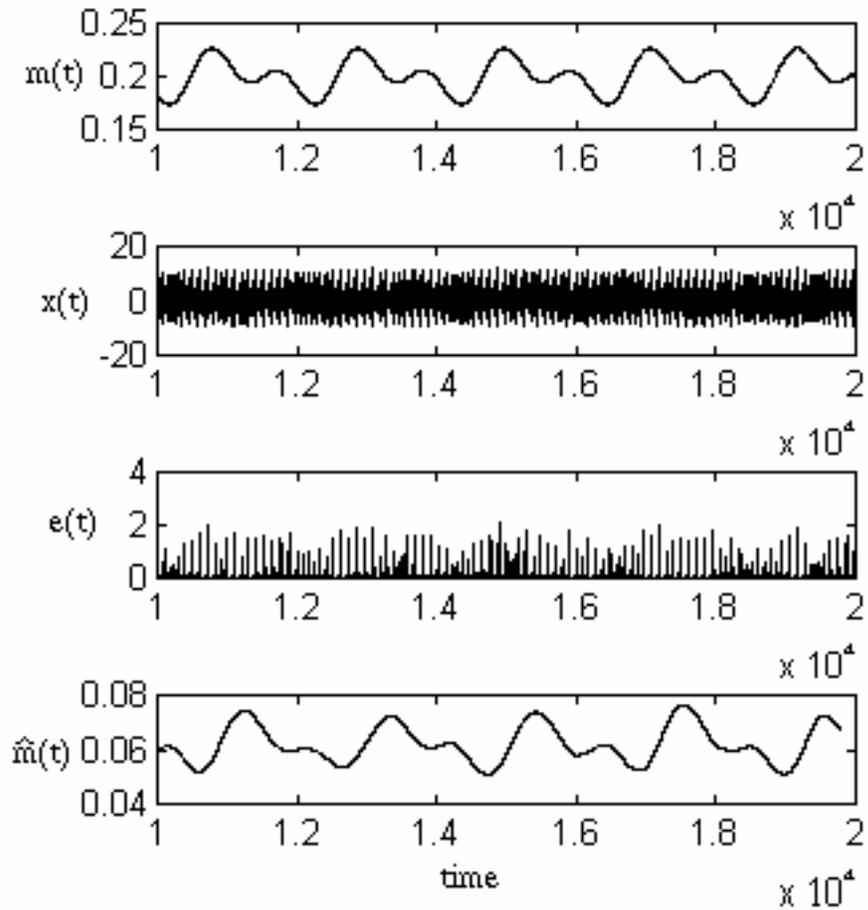}
  \caption{ Results for the Rossler oscillator. }
 \end{figure}

 Again, the fourier spectrum clearly shows
 the modulating waveforms. The variation of
the
 output or the $\lambda$'s may not be linear when modulating parameter
varies by a large
  amount. To compensate for this, a non-linear
companding/expanding circuitry
  may be used after demodulation.
 Thus, it is easy to intercept and decode a continous-time
 secure chaotic communication.

 In conclusion, we have presented a simple algorithm for decoding an unknown
 modulated chaotic time-series. This technique can be used for
 real-time decoding
 using a DSP or an analog circuitry.

 \end{document}